\documentclass[proceedings, preprint]{rmaa}

\usepackage{paralist}

\usepackage{psfrag,color}

\SetYear{2011}
\SetConfTitle{Atmospheric Data from Astronomical Site Testing in Chile}

\title{Possibility of Terahertz Observations at the ALMA site}

\author{Satoki Matsushita\altaffilmark{1,2}}

\altaffiltext{1}{Academia Sinica Institute of Astronomy and
	Astrophysics, P.O.\ Box 23-141, Taipei 10617, Taiwan,
	R.O.C.\ (satoki@asiaa.sinica.edu.tw).}

\altaffiltext{2}{Joint ALMA Office, Alonso de C\'ordova 3107,
	Vitacura 763 0355, Santiago, Chile.}

\shortauthor{Matsushita}
\shorttitle{Possibility of Terahertz Observations at the ALMA site}

\listofauthors{S. Matsushita}
\indexauthor{Matsushita, S.}

\abstract{
Observational rates under terahertz (THz) opacities less than 3.0 and
2.0 at the Atacama Large Millimeter/submillimeter Array (ALMA) site
have been calculated using the 225~GHz tipping radiometer monitoring
data and the opacity correlation between 225~GHz and THz opacities.
The observational rate with THz opacity condition less than 3.0
is 12.4\% in a year, and in winter (November -- April) it is about
twice higher than in summer (May -- October).
This observational rate shows a large sinusoidal annual variation,
and it seems to have relation with the El Ni\~no and La Ni\~na
phenomena; the La Ni\~na years tend to have high observational rates,
but the El Ni\~no years show low rates.
On the other hand, the observational rate with the THz opacity
condition less than 2.0 is only 1.9\%, and no obvious annual and
seasonal variations are observed.
This indicates that THz observations under low opacity condition of
less than 2.0 at the ALMA site are very difficult to be performed.
}

\resumen{
El porcentaje de observacion con opacidades menores que 3.0 y 2.0 en
frecuencias THz en el lugar de Atacama Large Millimeter/submillimeter
Array (ALMA), han sido calculadas usando el monitoreo de los datos
radiometricos a 225 GHz y la correlacion de opacidades entre 225 GHz
y el rango de frecuencias THz.
Este porcentaje de observacion con opacidades menores que 3.0 en
frecuencias THz se da el 12.4\% del a\~no, y durante el invierno
(Noviembre -- Abril), es cerca del dos veces mayor que en verano
(Mayo -- Octubre).
Este porcentaje de observacion muestra una larga variacion sinusoidal
a lo largo del a\~no, y parece estar relacionado con los fenomenos de
el Ni\~no y la Ni\~na; durante los anyos en los que sucede el
fenomenos de la Ni\~na el porcentaje de observacion tiende a ser mas
alto, pero en los a\~nos en que el fenomeno del Ni\~no se manifiesta,
el porcentaje decrece.
Por otro lado, el porcentaje de observacion con opacidades menores
que 2.0 en frecuencias THz es tan solo del 1.9\% y no muestra una
variacion anual obvia entre las diferentes estaciones.
Esto indica que las observaciones en THz bajo condiciones de baja
opacidad menor que 2.0 son muy complicadas de realizar en ALMA.
}

\addkeyword{Atmospheric effects}
\addkeyword{Site testing}
\addkeyword{Submillimeter: General}

\begin{document}
% Typeset article header
\maketitle

\section{Introduction}
\label{sect-intro}

At the terahertz (THz) frequency range, various emission/absorption
lines can be seen from astronomical sources; high transition
molecular lines (e.g., CO J = 11 -- 10 or higher), atomic lines
(e.g., [\ion{N}{II}] $205~\mu$m), and redshifted infrared lines
(e.g., [\ion{C}{II}] $158~\mu$m and [\ion{O}{I}] $63~\mu$m), which
are useful for the study of warm gas in various sources.
In addition, low temperature (a few 10~K) dust continuum emission
also peaks around this frequency range.
%, which is useful for the low temperature gas studies.
However, this frequency range is very difficult to be observed from
ground telescopes since the atmospheric absorption is very strong,
and therefore it is observable at very limited sites under very
limited weather conditions.
Going to space or upper atmosphere is another option, but it is
usually very expensive, so it is also not easy.
Here, in this paper, I present an estimation of the observable time
at the THz range at one of the best observation sites in the world,
the Atacama Large Millimeter/submillimeter Array (ALMA) site.

ALMA is composed by up to eighty high-precision antennas, located
at the Chajnantor plain of the Chilean Andes, near San Pedro de
Atacama, 5000~m above the sea level.
It is currently under construction and commissioning with the
collaboration between East Asia, Europe, and North America
\citep{hil10,woo09}.
Before starting the construction, various site testing measurements
had been done around this site, including the 220~GHz/225~GHz tipping
radiometer measurements \citep{koh95,rad98,rad00,rad01,rad02,sak02}
and the Fourier Transform Spectrometer (FTS) measurements
\citep{mat98,mat99,pai00,mat03}.
The former measured data provided the long-term (up to a decade)
220~GHz/225~GHz opacity variation, and the latter measured data
provided the atmospheric opacity spectra from millimeter to
submillimeter wavelengths, and even up to THz opacity.
In addition, some of the latter data also provided the correlation
between 220~GHz/225~GHz opacity and other opacities.
In this paper, I use the 225~GHz opacity data taken with the National
Radio Astronomy Observatory (NRAO) tipping radiometer and the opacity
correlation derived from the National Astronomical Observatory of
Japan (NAOJ) FTS opacity spectrum data, to estimate the time variation
of the THz opacity.

\section{Data Reduction}
\label{sect-data}

For the opacity time variation data of the NRAO 225~GHz tipping
radiometer data, I used data between April 1995 and April 2006
from the NRAO ALMA Site Characterization Homepage\footnote{
\url{http://science.nrao.edu/alma/site-characterization.shtml}} and
Simon Radford's Chajnantor Site Evaluation Homepage\footnote{
\url{http://www.submm.caltech.edu/~sradford/site-eval/}}.
I convert these 225~GHz opacity data into the THz opacity using the
opacity correlation between 225~GHz and THz derived from the NAOJ FTS
data:
The correlation coefficient between the 220~GHz opacity and the
opacities at 1035~GHz, 1350~GHz, and 1500~GHz atmospheric windows are
derived as $123\pm5$, $115\pm29$, and $105\pm32$, respectively
\citep{mat99}.
In this paper, I assume THz opacity is 105 times larger than 225~GHz
opacity, namely ${\rm [THz~opacity] = 105 \times [225~GHz~opacity]}$.

First, I multiply by 105 all the available 225~GHz opacity data to
derive the THz opacity, and then I calculate how much data points are
below the following two opacity conditions (i.e., observational
rates); one is the THz opacity, $\tau_{\rm THz}$, less than 3.0
($\tau_{\rm 225~GHz} < 0.029$), and another less than 2.0
($\tau_{\rm 225~GHz} < 0.019$).
These calculated values have been used to estimate the annual and
seasonal (summer and winter) variations of the observational rates.
Here, I assume summer as the time between November and April, and
winter between May and October.

\section{Results}
\label{sect-res}

Table~\ref{tab-sum} shows the observational rates under two opacity
conditions.
Under the condition of $\tau_{\rm THz} < 3.0$, it is possible to
observe 12.4\% of a year ($\sim45$ days).
Seasonal difference is obvious, 16.8\% of the time is observable at
winter, but only about a half (8.7\%) in summer.
On the other hand, under the condition of $\tau_{\rm THz} < 2.0$,
which is the best weather condition at the ALMA site, there is no
clear seasonal difference, and the observational rate is only 1.9\%
of a year, and 2.0\% and 1.7\% for winter and summer, respectively.

\begin{table}[!t]\centering
  \setlength{\tabnotewidth}{\columnwidth}
  \tablecols{3}
  % Stretch the space between table columns 
  %\setlength{\tabcolsep}{2.8\tabcolsep}
  \caption{Annual and Seasonal Observational Rates at Terahertz}
  \label{tab-sum}
  \begin{tabular}{lrr}
    \toprule
    & $\tau_{\rm THz} < 3.0$ & $\tau_{\rm THz} < 2.0$ \\
    \midrule
    Summer (Nov.~-~Apr.) &  8.7\% & 1.7\% \\
    Winter (May~-~Oct.)  & 16.8\% & 2.0\% \\
    Annual               & 12.4\% & 1.9\% \\
    \bottomrule
  \end{tabular}
\end{table}

\begin{figure}[!t]
  \includegraphics[width=\columnwidth]{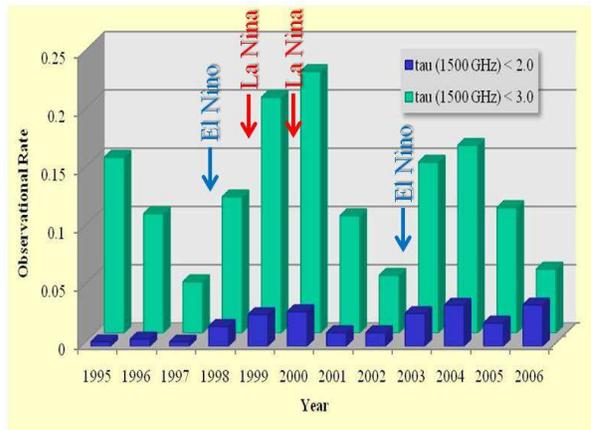}
  \caption{
    Annual variation of the observational rates under the conditions
    of $\tau_{\rm THz} < 3.0$ (light histogram in the back) and
    $\tau_{\rm THz} < 2.0$ (dark histogram in front) between 1995
    and 2006.}
  \label{fig-av1}
\end{figure}

Figure~\ref{fig-av1} depicts the annual variation of the
observational rates under the conditions of $\tau_{\rm THz} < 3.0$
(light histogram) and $\tau_{\rm THz} < 2.0$ (dark histogram) between
1995 and 2006.
The observational rate under the condition of $\tau_{\rm THz} < 3.0$
shows large annual variation, with the highest (22.4\% or 82 days on
2000) and the lowest (4.4\% or 16 days on 1997) rates differs for
almost 20\%.
In addition, it exhibits a sinusoidal variation within an epoch of
about 4 -- 5 years, with high observational rate on 1995, 1999/2000,
and 2003/2004, but low on 1997, 2002, and 2006.
On the other hand, the observational rate under the condition of
$\tau_{\rm THz} < 2.0$ shows small annual variation, with the highest
(3.5\% or 13 days on 2004 and 2006) and the lowest (0.36\% or 1 day
on 1997) rates differing only a few \%.

\begin{figure}[!t]
  \includegraphics[width=\columnwidth]{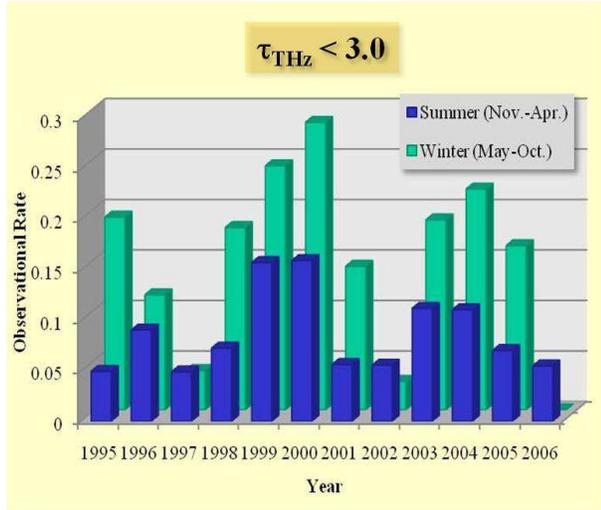}
  \caption{
    Annual variation of the observational rates in winter (light
    histogram in the back) and summer (dark histogram in front) under
    the conditions of $\tau_{\rm THz} < 3.0$ between 1995 and 2006.}
  \label{fig-av2}
\end{figure}

Figure~\ref{fig-av2} depicts the annual variations of winter (light
histogram) and summer (dark histogram) under the opacity condition of
$\tau_{\rm THz} < 3.0$.
The annual variation of the winter season follows well the sinusoidal
trend of the annual variation mentioned above, which is high
observation rate on 1995, 2000, and 2004, but low on 1997, 2002, and
2006.
In addition, the difference between the high and low observation rate
is more pronounced than that of the annual variation mentioned above;
the highest rate is 28.5\% (or 52 days) on 2000, but the lowest rate
is only 2.8\% (or 5 days) on 2002, with a difference of more than
25\% (note that there is no data in the winter season of 2006).
On the other hand, the summer season shows less variation than that
of the winter season, with less pronounced sinusoidal variation.
It is interesting to note that in the year of the low observational
rate (1997 and 2002), there is almost no difference in the
observational rates between winter and summer (actually, the rate in
summer is higher than that in winter, although it is very small
difference).

\begin{figure}[!t]
  \includegraphics[width=\columnwidth]{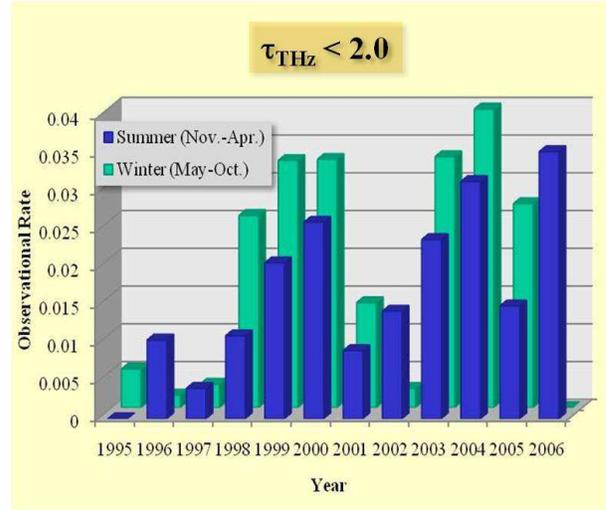}
  \caption{
    Annual variation of the observational rates in winter (light
    histogram in the back) and summer (dark histogram in front) under
    the conditions of $\tau_{\rm THz} < 2.0$ between 1995 and 2006.}
  \label{fig-av3}
\end{figure}

Figure~\ref{fig-av3} exhibits the annual variations of winter (light
histogram) and summer (dark histogram) under the opacity condition of
$\tau_{\rm THz} < 2.0$.
Compared to the annual variation of the condition of
$\tau_{\rm THz} < 3.0$, that of the condition of
$\tau_{\rm THz} < 2.0$ shows less seasonal variation, and sometimes
it shows good observational rate in summer than in winter.
The sinusoidal annual variation is also less pronounced.

\section{Disucssion}
\label{sect-dis}

The observational rate under the opacity condition of
$\tau_{\rm THz} < 3.0$ ($\tau_{\rm 225~GHz} < 0.029$) clearly shows
that there are large seasonal variations, suggesting that
observations under this opacity condition should be done in winter.
The observational rate under this opacity condition also exhibits
large annual variation.
I compared the occurrence of El Ni\~no and La Ni\~na with this annual
variation.
The El Ni\~no and La Ni\~na phenomena are derived from the Oceanic
Ni\~no Index (ONI) table in the homepage of the National Oceanic and
Atmosphere Administration (NOAA).\footnote{
\url{http://www.cpc.ncep.noaa.gov/products/analysis_monitoring/ensostuff/ensoyears.shtml}}
%taken from a homepage ``El Ni\~no and La Ni\~na Years and
%Intensities'' written by Jan Null\footnote{
%\url{http://ggweather.com/enso/oni.htm}}, and the events are
%calculated based on Oceanic Ni\~no Index.
From this comparison, it is found that the observational rate under
$\tau_{\rm THz} < 3.0$ is clearly bad at El Ni\~no years (1997 and
2002), but obviously very good at La Ni\~na years (1999 and 2000; see
Figure~\ref{fig-av1}).
This clearly suggests that El Ni\~no and La Ni\~na phenomena affect
the THz (and also millimeter and submillimeter) observational rates
at the ALMA site.

On the other hand, the observational rate under the best opacity
condition at the ALMA site, namely $\tau_{\rm THz} < 2.0$
($\tau_{\rm 225~GHz} < 0.019$), exhibits less annual and seasonal
variations with very rare observational occasion of only about 2\%
per year.
This indicates that very good opacity conditions are very rare
during all the year, and it is not very recommended to target these
weather conditions at the ALMA site.

\end{document}